# Impact Of Squark Generation Mixing On The Search For Gluinos At LHC

K. Hidaka

*Department of Physics, Tokyo Gakugei University, Koganei, Tokyo 184-8501, Japan*

**Abstract.** We study the effect of squark-generation mixing on gluino decays in the Minimal Supersymmetric Standard Model (MSSM). We show that due to the effect the quark-flavor violating (QFV) gluino decay branching ratio $B(\tilde{g} \to c\bar{t}(\bar{c}t)\tilde{\chi}_1^0)$ can be very large (up to ~50%) in a significant part of the MSSM parameter space despite the very strong experimental constraints on QFV from B meson observables. This could have an important impact on the search for gluinos and the determination of the MSSM parameters at LHC.

## INTRODUCTION

The decays of gluinos and squarks are usually assumed to be quark-flavour conserving. However, the squarks are not necessarily quark-flavor eigenstates and they are in general mixed. In this case quark-flavour violating (QFV) decays of gluinos and squarks could occur. The effect of QFV in the squark sector on reactions at colliders has been studied only in a few publications [1]. In this article based on [2] we study the effect of QFV due to the mixing of charm-squark and top-squark on the gluino and squark decays in the most general Minimal Supersymmetric Standard Model (MSSM).

## SQUARK MIXING WITH FLAVOR VIOLATION

The most general up-type squark mass matrix including left-right mixing as well as quark-flavor mixing in the super-CKM basis of $\tilde{u}_{0\gamma} = (\tilde{u}_L, \tilde{c}_L, \tilde{t}_L, \tilde{u}_R, \tilde{c}_R, \tilde{t}_R)$, $\gamma = 1,...,6$ is [2]

$$M_{\tilde{u}}^2 = \begin{pmatrix} M_{\tilde{u}LL}^2 & M_{\tilde{u}LR}^2 \\ M_{\tilde{u}RL}^2 & M_{\tilde{u}RR}^2 \end{pmatrix}, \qquad (1)$$

$$\left(M_{\tilde{u}LL}^2\right)_{\alpha\beta} = M_{Q_u\alpha\beta}^2 + \left[(\tfrac{1}{2} - \tfrac{2}{3}\sin^2\theta_W)\cos 2\beta\, m_Z^2 + m_{u_\alpha}^2\right]\delta_{\alpha\beta}, \qquad (2)$$

$$\left(M_{\tilde{u}RR}^2\right)_{\alpha\beta} = M_{U\alpha\beta}^2 + \left[\tfrac{2}{3}\sin^2\theta_W \cos 2\beta\, m_Z^2 + m_{u_\alpha}^2\right]\delta_{\alpha\beta}, \qquad (3)$$

$$\left(M_{\tilde{u}RL}^2\right)_{\alpha\beta} = (v_2/\sqrt{2})A_{U\beta\alpha} - m_{u_\alpha}\mu^*\cot\beta\,\delta_{\alpha\beta}. \qquad (4)$$

$M_{\tilde{u}LR}^2$ is the hermitean conjugate of $M_{\tilde{u}RL}^2$. The indices α,β=1,2,3 characterize the

quark flavors u,c,t, respectively. $M_{Q_u}^2$ and $M_U^2$ are the hermitean soft-supersymmetry (SUSY)-breaking mass matrices for the left and right up-type squarks, respectively. Note that $M_{Q_u}^2 = K \cdot M_Q^2 \cdot K^{-1}$, where $M_Q^2$ is the soft-SUSY-breaking mass matrix for the left down-type squarks and $K(\approx 1)$ is the CKM matrix. $m_{u_\alpha}$ ($u_\alpha = u,c,t$) are the physical quark masses. The physical mass eigenstates $\tilde{u}_i$, $i = 1,\ldots,6$, are given by $\tilde{u}_i = R_{i\alpha}^{\tilde{u}} \tilde{u}_{0\alpha}$. The mixing matrix $R^{\tilde{u}}$ and the mass eigenvalues are obtained by an unitary transformation $R^{\tilde{u}} M_{\tilde{u}}^2 R^{\tilde{u}-1} = diag(m_{\tilde{u}_1},\ldots,m_{\tilde{u}_6})$, where $m_{\tilde{u}_i} < m_{\tilde{u}_j}$ for $i < j$. We define the QFV parameters $\delta_{\alpha\beta}^{uLL}$ and $\delta_{\alpha\beta}^{uRR}$ ($\alpha \neq \beta$) as follows:

$$\delta_{\alpha\beta}^{uLL} \equiv M_{Q\alpha\beta}^2 / \sqrt{M_{Q\alpha\alpha}^2 M_{Q\beta\beta}^2} \tag{5}$$

$$\delta_{\alpha\beta}^{uRR} \equiv M_{U\alpha\beta}^2 / \sqrt{M_{U\alpha\alpha}^2 M_{U\beta\beta}^2} . \tag{6}$$

The down-type squark mass matrix can be analogously parametrized as the up-type squark mass matrix. As $M_Q^2 \approx M_{Q_u}^2$, one has $(M_{\tilde{d}LL}^2)_{\alpha\beta} \approx (M_{\tilde{u}LL}^2)_{\alpha\beta}$ for $\alpha \neq \beta$. We do not introduce additional QFV terms in the down-type squark mass matrix.

The properties of the charginos $\tilde{\chi}_i^\pm$ ($i=1,2, m_{\tilde{\chi}_1^\pm} < m_{\tilde{\chi}_2^\pm}$) and neutralinos $\tilde{\chi}_k^0$ ($k=1,\ldots,4$, $m_{\tilde{\chi}_1^0} < \ldots < m_{\tilde{\chi}_4^0}$) are determined by the parameters $M_2, M_1, \mu$ and $\tan\beta$, where $M_2$ and $M_1$ are the SU(2) and U(1) gaugino masses, respectively. Assuming gaugino mass unification including the gluino mass $m_{\tilde{g}} = M_3$, we take $M_1 = (5/3)\tan^2\theta_W M_2$.

## CONSTRAINTS

We impose the following conditions on the MSSM parameter space in order to respect experimental and theoretical constraints which are described in detail in [2]:
(i) Constraints from the B-physics experiments [3], such as
$3.03 \cdot 10^{-4} < B(b \to s\gamma) < 4.01 \cdot 10^{-4}$, $|\Delta M_{B_s}^{SUSY} - 17.77| < 3.31 ps^{-1}$ and so on.
(ii) The experimental limit on SUSY contributions to the electroweak $\rho$ parameter.
(iii) The LEP and Tevatron limits on the SUSY particle masses [3].
(iv) Vacuum stability conditions for the trilinear couplings $A_{U\alpha\beta}$ and $A_{D\alpha\beta}$ [3]. Conditions (i) and (iv) strongly constrain the 2nd and 3rd generation squark mixing parameters $M_{Q23}^2$, $M_{D23}^2$, $A_{U23}$, $A_{D23}$ and $A_{D32}$.

## QUARK FLAVOUR VIOLATING GLUINO DECAYS

Possible two-body decay modes of the gluino and squarks are:
$\tilde{g} \to \tilde{u}_i u_k$, $\tilde{d}_i d_k$

$$\tilde{u}_i \to u_k \tilde{\chi}_n^0, \, d_k \tilde{\chi}_m^+, \, \tilde{d}_j W^+, \, \tilde{u}_j Z^0, \, \tilde{u}_j h^0, \tag{7}$$

where $u_k = u,c,t$ and $d_k = d,s,b$. $h^0$ is the lighter CP-even Higgs boson. The squark decays into the heavier Higgs bosons, such as the CP-odd Higgs $A^0$, are kinematically forbidden in our scenario studied below. We take $\tan\beta$, $m_{A^0}$, $M_{1,2}$, $m_{\tilde{g}}$, $\mu$, $M^2_{Q\alpha\beta}$, $M^2_{U\alpha\beta}$, $M^2_{D\alpha\beta}$, $A_{U\alpha\beta}$ and $A_{D\alpha\beta}$ as the basic MSSM parameters at the weak scale. We assume them to be real. The QFV parameters are $M^2_{Q\alpha\beta}$, $M^2_{U\alpha\beta}$, $M^2_{D\alpha\beta}$, $A_{U\alpha\beta}$ and $A_{D\alpha\beta}$ with $\alpha \neq \beta$. We take the following scenario as a reference scenario with QFV within the reach of LHC (All mass parameters are in GeV.):

$\tan\beta = 10$, $m_{A^0} = 800$, $M_1 = 139$, $M_2 = 264$, $m_{\tilde{g}} = 800$, $\mu = 1000$, $M^2_{Q11} = (920)^2$, $M^2_{Q22} = (880)^2$, $M^2_{Q33} = (840)^2$, $M^2_{Q12} = M^2_{Q13} = 0$, $M^2_{Q23} = (224)^2$, $M^2_{U11} = (820)^2$, $M^2_{U22} = (600)^2$, $M^2_{U33} = (580)^2$, $M^2_{U12} = M^2_{U13} = 0$, $M^2_{U23} = (224)^2$, $M^2_{D11} = (830)^2$, $M^2_{D22} = (820)^2$, $M^2_{D33} = (810)^2$, $M^2_{D12} = M^2_{D13} = M^2_{D23} = 0$, and all of $A_{U\alpha\beta}$ and $A_{D\alpha\beta}$ are set to zero. In this scenario satisfying all the conditions (i)-(iv) above we have (Masses are in GeV.):

$$m_{\tilde{u}_1} \cong 558, \, m_{\tilde{u}_2} \cong 642, \, \tilde{u}_1 \cong 0.728 \tilde{c}_R - 0.685 \tilde{t}_R, \, \tilde{u}_2 \cong -0.686 \tilde{c}_R - 0.727 \tilde{t}_R, \, m_{\tilde{\chi}_1^0} = 138, \tag{8}$$

$$B(\tilde{g} \to ct\tilde{\chi}_1^0) = \sum_{i=1,2} \left[ B(\tilde{g} \to \tilde{u}_i c) B(\tilde{u}_i \to t\tilde{\chi}_1^0) + B(\tilde{g} \to \tilde{u}_i t) B(\tilde{u}_i \to c\tilde{\chi}_1^0) \right] = 0.463, \tag{9}$$

$$B(\tilde{g} \to cc\tilde{\chi}_1^0) = \sum_{i=1,2} \left[ B(\tilde{g} \to \tilde{u}_i c) B(\tilde{u}_i \to c\tilde{\chi}_1^0) \right] = 0.380, \tag{10}$$

$$B(\tilde{g} \to tt\tilde{\chi}_1^0) = \sum_{i=1,2} \left[ B(\tilde{g} \to \tilde{u}_i t) B(\tilde{u}_i \to t\tilde{\chi}_1^0) \right] = 0.120. \tag{11}$$

As we always sum over the particles and antiparticles of the (s)quarks, we do not indicate if it is a particle or its antiparticle: $B(\tilde{g} \to ct\tilde{\chi}_1^0) \equiv B(\tilde{g} \to c\bar{t}\tilde{\chi}_1^0) + B(\tilde{g} \to \bar{c}t\tilde{\chi}_1^0)$. Note that the QFV gluino decay branching ratio of Eq. 9 is very large. The reason is as follows: The gluino decays into squarks other than $\tilde{u}_{1,2}$ are kinematically forbidden, and $\tilde{u}_1$, $\tilde{u}_2$ are strong mixtures of $\tilde{c}_R$ and $\tilde{t}_R$ due to the large $\tilde{c}_R$-$\tilde{t}_R$ mixing term $M^2_{U23}$ ($=(224 \text{ GeV})^2$) in this scenario. This results in the very large QFV decay branching ratio $B(\tilde{g} \to ct\tilde{\chi}_1^0)$. Note that $\tilde{u}_{1,2}(\sim \tilde{c}_R + \tilde{t}_R)$ couple to $\tilde{\chi}_1^0 (\cong \tilde{B}^0)$ and practically do not couple to $\tilde{\chi}_2^0 (\cong \tilde{W}^0)$, $\tilde{\chi}_1^\pm (\cong \tilde{W}^\pm)$, and that $\tilde{\chi}_{3,4}^0$, $\tilde{\chi}_2^\pm$ are very heavy in this scenario. Here $\tilde{B}^0$ and $\tilde{W}^{0,\pm}$ are the U(1) and SU(2) gauginos, respectively.

In Fig.1 we show contours of $B(\tilde{g} \to ct\tilde{\chi}_1^0)$ in the $\delta_{23}^{uLL}$-$\delta_{23}^{uRR}$ plane where all of the conditions (i)-(iv) except the $b \to s\gamma$ constraint are satisfied. All basic parameters other than $M^2_{Q23}$ and $M^2_{U23}$ are fixed as in our reference scenario defined above. We see that $B(\tilde{g} \to ct\tilde{\chi}_1^0)$ increases quickly with increase of the $\tilde{c}_R$-$\tilde{t}_R$ mixing parameter $|\delta_{23}^{uRR}|$ and can be very large (up to ~ 50 %) in a significant part of the $\delta_{23}^{uLL}$-$\delta_{23}^{uRR}$ plane allowed by all of the conditions (i)-(iv) including the $b \to s\gamma$ constraint.

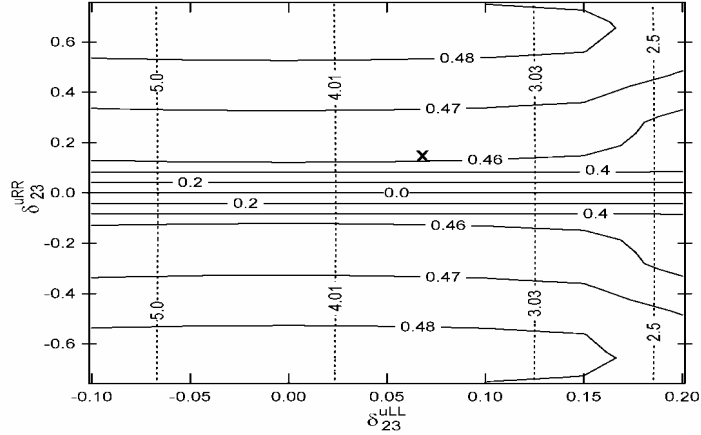

**FIGURE 1.** Contours of the QFV decay branching ratio $B(\tilde{g} \to ct\tilde{\chi}_1^0)$ (solid lines) in the $\delta_{23}^{uLL}$ - $\delta_{23}^{uRR}$ plane where all of the conditions (i)-(iv) except the $b \to s\gamma$ constraint are satisfied. Contours of $10^4 \cdot B(b \to s\gamma)$ (dashed lines) are also shown. The condition (i) requires $3.03 < 10^4 \cdot B(b \to s\gamma) < 4.01$. The point "x" of ($\delta_{23}^{uLL}, \delta_{23}^{uRR}$) = (0.068, 0.144) corresponds to our reference scenario.

The other QFV parameter dependences of $B(\tilde{g} \to ct\tilde{\chi}_1^0)$ are also shown in [2].

The signature of the decay $\tilde{g} \to ct\tilde{\chi}_1^0$ would be '(charm-) jet + top-quark + missing-energy'. In any case, one should take into account the possibility of significant contributions from QFV decays in the gluino search. One should also include the QFV squark parameters in the determination of the basic SUSY parameters at LHC.

## CONCLUSION

We have studied gluino decays in the MSSM with squark mixing of the second and third generation, especially $\tilde{c}_{L/R} - \tilde{t}_{L/R}$ mixing. We have shown that QFV gluino decay branching ratios such as $B(\tilde{g} \to ct\tilde{\chi}_1^0)$ can be very large due to the squark mixing in a significant part of the MSSM parameter space despite the very strong experimental constraints from B factories, Tevatron and LEP. This could have an important impact on the search for gluinos and the MSSM parameter determination at LHC.

## ACKNOWLEDGMENTS

I deeply thank the other authors of [2]: A. Bartl, K. Hohenwarter-Sodek, T. Kernreiter, W. Majerotto and W. Porod.